\documentclass[twocolumn,useAMS]{mn2e}
\usepackage{graphicx}
\usepackage{url}

\usepackage{bm}
\usepackage{amsmath}
\usepackage{amssymb}
\usepackage{epsf}

\usepackage{color} 
\newcommand{\bea}{\begin{eqnarray}}
\newcommand{\eea}{\end{eqnarray}}

\newcommand{\Mvir}{M_{\rm vir}}

\newcommand{\barsvar}{\overline{\mathcal{S}^2}}

\newcommand{\Sobs}{S_{\rm obs}}

\newcommand{\sobsbar}{\overline{\mathcal{S}}_{\rm obs}}
\newcommand{\svar}{\overline{\mathcal{S}^2}_{\rm obs}}
\newcommand{\snobs}{\overline{\mathcal{S}^n}_{\rm obs}}

\newcommand{\nobs}{n_{\rm obs}}
\newcommand{\Nobs}{N_{\rm obs}}
\newcommand{\dOmegaobs}{d\Omega_{\rm obs}}
\newcommand{\Omegaobs}{\Omega_{\rm obs}}
\newcommand{\Omegavir}{\Omega_{\rm vir}}

\newcommand{\Msun}{M_{\odot}}

\newcommand{\simgt}{\lower.5ex\hbox{$\; \buildrel > \over \sim \;$}}
\newcommand{\simlt}{\lower.5ex\hbox{$\; \buildrel < \over \sim \;$}}

\title{Magnification Effects on Source Counts and Fluxes} 
\author[Jain and Lima]{
Bhuvnesh Jain and Marcos Lima
\\
Department of Physics \& Astronomy, University of Pennsylvania, Philadelphia, PA 19104 \\
}
\begin{document}
\maketitle
\begin{abstract}
We consider the effect of lensing magnification on 
high redshift sources in the case that magnification varies on
the sky, as expected in wide fields of view or within  observed galaxy
clusters. We give expressions for number counts, flux
and flux variance as integrals over the probability distribution of
the magnification. We obtain these through a simple mapping between
averages over the observed sky and over the 
magnification probability distribution in the source plane. 
Our results clarify conflicting expressions in the literature and can 
be used to calculate a variety of magnification effects. We highlight two  
applications: 1. Lensing of high-$z$ galaxies by galaxy clusters 
can provide the dominant source of scatter in SZ observations at frequencies 
larger than the SZ null. 2. The number counts of high-$z$ galaxies with a 
Schechter-like luminosity function will be changed at high
luminosities to a power law, with significant enhancement of the
observed counts at $L\simgt 10\ L^*$. 
\end{abstract}

\vspace{0.9in}

\section{Introduction}

Magnification due to gravitational lensing leads to observable effects, 
namely changes in the number density of 
galaxies behind large-scale structure and galaxy clusters (known as
magnification bias) and  in the moments of the flux distribution due
to unresolved sources at  high redshift. 
These and other effects of lensing magnification 
have been studied extensively in the last few decades, usually
assuming simple expressions that apply for constant magnifications.

In this brief note, we generalize to the case where magnification
varies on the sky -- the variation is taken to be given by a
magnification probability in the  
{\it source plane}, while quantities of interested are observed as 
averages in the {\it image plane}. We apply this calculation to lensing of 
the intrinsic number count distributions of high redshift galaxies as well 
as moments of the flux for Poisson distributed high-$z$ galaxies 
behind galaxy clusters. 
Our goal is to provide the formulae needed for magnification effects
in a variety of physical situations and give estimates of the scale of the 
main effects. 
Applications to more detailed models and results for Sunyaev-Zel'dovich 
surveys have been presented in a separate paper (Lima, Jain \& Devlin
2009).  

\section{Constant Magnification}

By definition, magnification (denoted $\mu$) is the Jacobian of the 
transformation between image (lensed) and source (unlensed) coordinates
(e.g. Bartelmann \& Schneider 2001). 
Along a given line of sight, its effect on differential solid 
angles is given by 
\begin{eqnarray}
d\Omega&\rightarrow& d\Omegaobs=\mu d\Omega \, ,
\end{eqnarray}
or $\mu=d\Omegaobs/d\Omega$.  We use subscript ``$\rm obs$'' for the observed (or
lens plane or image plane) and no subscript for the (unlensed)  
source plane. The surface brightness of 
galaxy sources, defined as the flux per unit solid angle, is conserved by 
lensing.
Since magnification increases the solid angle 
of sources by a factor $\mu$, it also increases their flux $S$ as 
\begin{eqnarray}
S&\rightarrow& \Sobs=\mu S\, . 
\end{eqnarray}
In terms of the lensing shear $\gamma$ and
convergence $\kappa$, the magnification is given by 
$\mu = 1/[(1-\kappa)^2 - |\gamma|^2]$. 

As a result, the number density of a source population 
is modified by lensing magnification. Let $dn/dS$ denote the intrinsic number density 
per unit flux per unit steradian on the sky. 
Given  a (constant) magnification $\mu$, it is modified as 
\begin{eqnarray}
\frac{dn}{dS} \rightarrow \frac{d\nobs(\Sobs)}{d\Sobs} = 
\frac{1}{\mu^2} \frac{dn}{dS}\left(\frac{\Sobs}{\mu}\right) \,.
\end{eqnarray}
The $1/\mu^2$ factor comes from transforming the angle $d \Omega$ 
and the flux differential $dS$ into their observed counterparts using Eqns. 1 
and 2. 
The change in argument comes from 
the fact that the observed flux $\Sobs$ corresponds to true flux 
$S=\Sobs/\mu$. 

Given the differential number density $dn/dS$, we may define the cumulative 
number density $n(>S)$, the average flux of the background 
galaxy population per steradian 
$\overline{\mathcal{S}} $ and the mean square flux per steradian 
$\overline{\mathcal{S}^2} $ as 
\begin{eqnarray}
n(>S)&=&\int_S \frac{dn}{dS^\prime} dS^\prime \,,\\
\overline{\mathcal{S}} &=& \int S \frac{dn}{dS} dS \,,\\
\overline{\mathcal{S}^2} &=& \int S^2 \frac{dn}{dS}dS\,.
\label{intrinsic}
\end{eqnarray}
In the presence of a constant mangification $\mu$, the observed
quantities are easily obtained using Eqns. 2 and 3 as: 
\begin{eqnarray}
n(>S)\rightarrow \frac{1}{\mu} n\left(>\frac{\Sobs}{\mu}\right), \ \ \
\overline{\mathcal{S}}\rightarrow \overline{\mathcal{S}},  \ \ \  
\overline{\mathcal{S}^2}\rightarrow  \mu\ \overline{\mathcal{S}^2}. 
\end{eqnarray}
Note that in the integrals over $S$ for $\overline{\mathcal{S}}$ and
$\overline{\mathcal{S}^2}$  there is no upper or lower cutoff in flux. 

There is a long history in
the literature of magnification effects on source 
counts 
(starting with Canizares 1981, 1982 and Peacock 1982).  
The expressions above 
are consistent with those in the literature. 
We next consider the case 
of variable magnification on the sky. 

\section{Variable magnification on the sky}

\begin{figure*}
\resizebox{85mm}{!}{\includegraphics[angle=0]{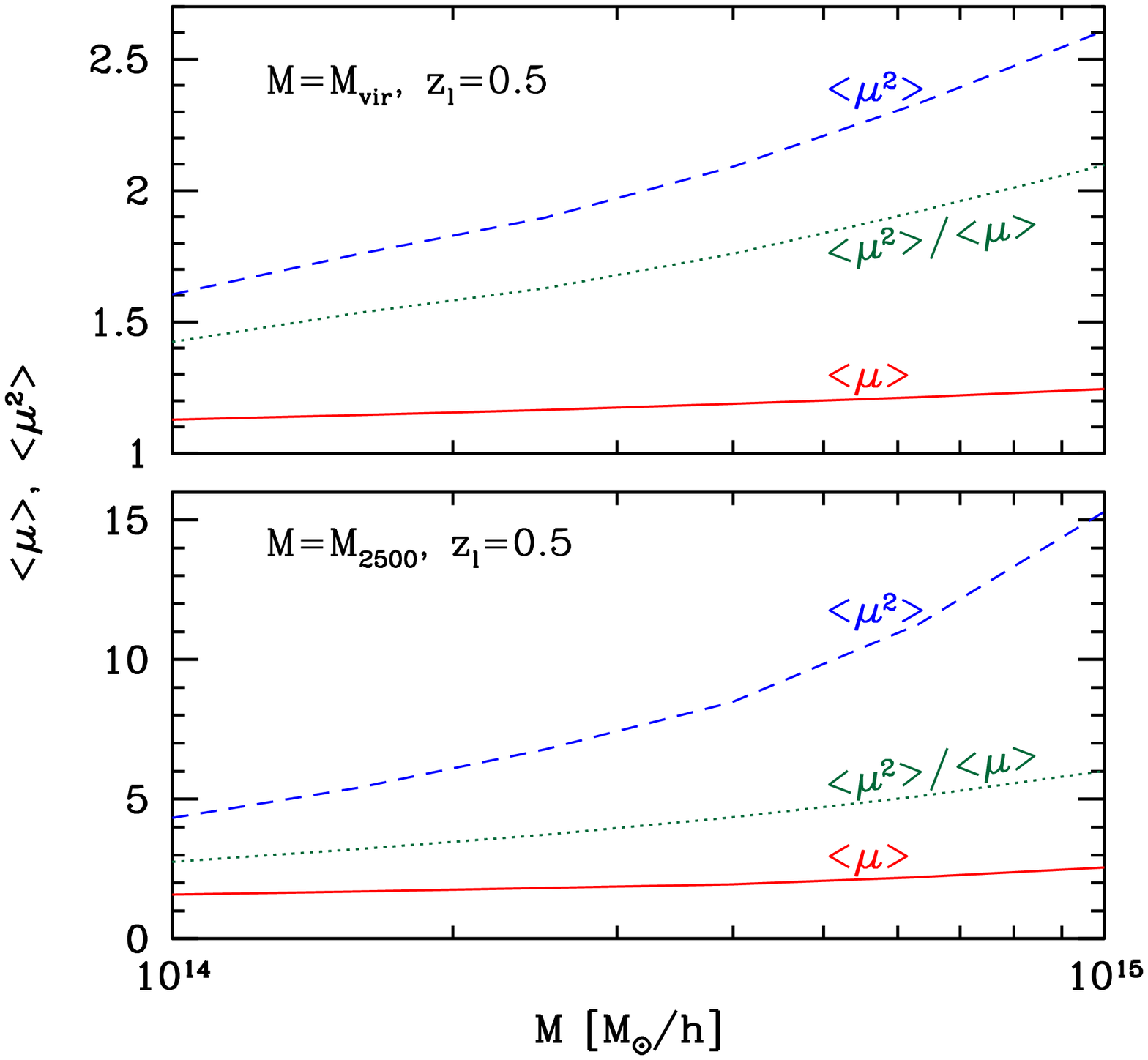}}
\resizebox{85mm}{!}{\includegraphics[angle=0]{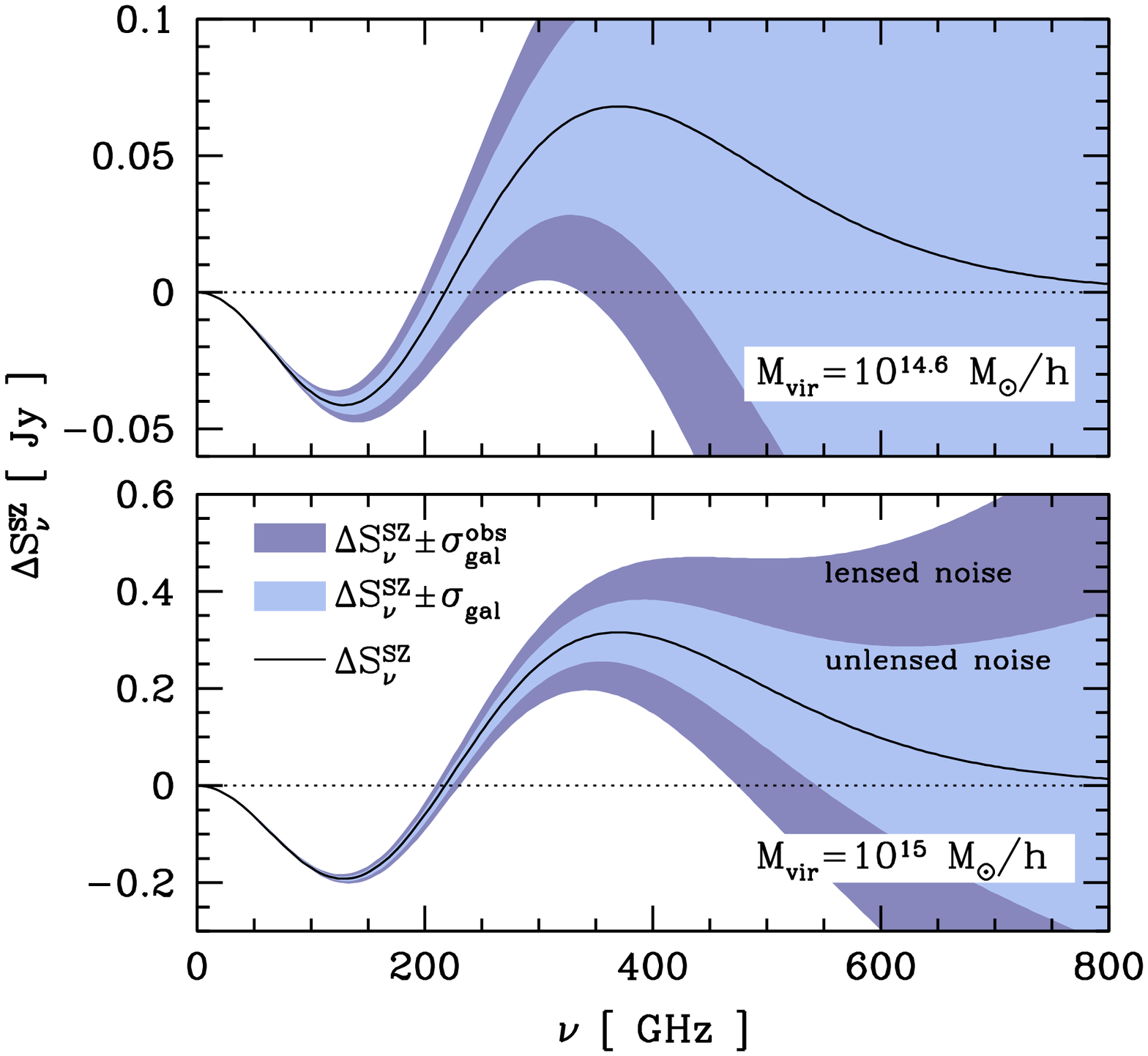}}
  \caption{{\it Left:} Mean magnification $\langle \mu \rangle$ and mean squared
$\langle \mu^2 \rangle$ measured within clusters at redshift $z=0.5$ 
as a function of cluster mass. The upper panel averages within the 
virial radius of the cluster while the lower panel uses a smaller radius 
within which the overdensity is 2500 times the critical density. 
In X-ray and SZ analysis, different choices for the cluster radius are 
made which typically lie within these two. 
{\it Right:} Contamination of submillimeter galaxies to the SZ flux
within the virial radius of clusters of virial mass $\Mvir=10^{14.6}$ and 
$10^{15}h^{-1}\Msun$. The solid line shows the intrinsic SZ flux 
$\Delta S_{\nu}^{\rm SZ}$. The light shaded region delineate 
the 1-$\sigma$ region from contamination of submillimeter galaxies 
from Poisson noise in their counts, i.e. 
$\sigma_{\rm gal}^2=\overline{\mathcal{S}^2}_{\rm vir}$.
The dark shaded region accounts for lensing magnification, which further
enhances the noise by $\langle \mu^2 \rangle/\langle \mu \rangle$, cf. Eqn~22.
}
\label{fig:avemu}
\end{figure*}

We wish to generalize Eqns. 3 and 7 to the case that the magnification 
varies on the sky. This variation can occur over a large patch of the
sky with fluctuations due to large-scale structure or simply 
over the surface of a galaxy cluster due to variations in the surface
mass density and shear over this surface. 

For variable magnification, the obvious step would be to 
average Eqns. 3 and 7 over the observed sky
(i.e. the image plane), and this is indeed correct. Thus 
Eqn. 3 generalizes to 
$ d\nobs/d\Sobs = \int d\mu \ dn/dS \ P_{\rm obs}(\mu)/\mu^2 $, 
where $P_{\rm obs}(\mu)$ is the normalized magnification probability in
the {\it image} plane. It is
often preferable to do calculations in the {\it source} plane. 
This is straightforwardly done by using the relation
\begin{equation} 
P_{\rm obs} (\mu) = \frac{\mu}{\langle\mu\rangle} \ P(\mu)
\end{equation}
where $P(\mu)$ is the {\it source} plane probability. 

The above relation gives the generalized expressions: 
\begin{eqnarray}
\left\langle\frac{d\nobs(\Sobs)}{d\Sobs}\right\rangle &=& 
\frac{1}{\langle\mu\rangle}\int d\mu \frac{P(\mu)}{\mu}\
\frac{dn}{dS}\left(\frac{\Sobs}{\mu}\right)\,,  
\\ 
\langle\nobs(>\Sobs)\rangle &=& 
\frac{1}{\langle\mu\rangle}\int d\mu P(\mu)\  n\left(>\frac{\Sobs}{\mu}\right) \,, \\ 
\langle{\sobsbar}\rangle &=& \overline{\mathcal{S}}\,, \\
\langle\svar\rangle &=& \frac{\langle\mu^2\rangle}{\langle\mu\rangle} 
\overline{\mathcal{S}^2}\, , 
\label{lensed} 
\end{eqnarray}
where $\langle \rangle$ denote averages of observed quantities over  
specified parts of the sky. 

To obtain these results more formally, 
we evaluate the expressions on the LHS of the above equations by 
defining the average of a function $X$ in the image plane over an 
{\it observed} solid angle as 
\begin{equation}
\langle X \rangle_{\rm obs} \equiv \frac{1}{\Delta\Omegaobs}\int \dOmegaobs X\,.
\end{equation}
In the source (unlensed) plane the average over solid angle also 
{\it defines} $P(\mu)$
\begin{equation}
\langle X \rangle_{\rm source} \equiv \frac{1}{\Delta\Omega}\int d\Omega\ X \equiv
\int d\mu P(\mu)\ X\,.
\end{equation}
The function $X$ is a function of angle 
${\bf\theta}$ on the sky 
through its dependence on $\mu({\bf\theta})$. 
Defining $P(\mu)$ on the source plane is conventional in lensing
as it addresses questions such as, what is the fraction of sources
that are magnified by a certain amount? 
\footnote{One must of course ensure that
theoretical predictions are also in the source plane. This occurs naturally in ray
tracing simulations which start the rays at the observer and trace
backwards (e.g. Jain, Seljak \& White 2000). However predictions that
rely on the Born approximation apply to the image plane. 
See Hilbert et al. (2008) for a discussion of
simulation predictions; they also consider the effects of multiple
imaging which we are not concerned with here. 
}
Note that we have as desired
\begin{eqnarray}
\int d\mu P(\mu) &=& \frac{1}{\Delta\Omega}\int d\Omega\ =1\,,\\ 
\int d\mu\ \mu P(\mu) &=&  \frac{1}{\Delta\Omega}\int d\Omega\ \mu(\theta)
\equiv \langle \mu \rangle =\frac{\Delta\Omegaobs}{\Delta\Omega}\,.
\label{meanmu}
\end{eqnarray}
In the limit of the whole sky, we have $\Delta\Omegaobs=\Delta\Omega=4\pi$ and 
$\langle \mu \rangle=1$, i.e. the average magnification is unity. 

Using the relations given above in Eqns. 13-16 for angular averaging,
we can  obtain Eqn. 9 as follows:  
\begin{eqnarray}
\left\langle\frac{d\nobs(\Sobs)}{d\Sobs}\right\rangle &=&  
\frac{1}{\Delta\Omegaobs}\int \dOmegaobs
\frac{d\nobs(\Sobs)}{d\Sobs}\\ \nonumber
&=&  \frac{1}{\langle\mu\rangle \Delta\Omega}\int d\Omega\ \mu
\frac{1}{\mu^2} \frac{dn}{dS}\left(\frac{\Sobs}{\mu}\right) \\ \nonumber
&=&\frac{1}{\langle\mu\rangle} \int d\mu\ P(\mu) \frac{1}{\mu} 
\frac{dn}{dS}\left(\frac{\Sobs}{\mu}\right)\,.\nonumber
\end{eqnarray}
This is our first desired result. It is obviously different from integrating
the expression for $d\nobs/d\Sobs$ from Eqn. 3
over $P(\mu)$ -- doing that would have led to both factors of $1/\mu$
being inside the integrand. 
Next we substitute 
Eqn. 3 into
\begin{eqnarray}
\langle\nobs(>\Sobs)\rangle = 
\frac{1}{\Delta\Omegaobs}\int \dOmegaobs
\int_{\Sobs} \frac{d\nobs(\Sobs')}{d\Sobs'} d\Sobs'\,, \nonumber
\end{eqnarray}  
and change variables to $S'=\Sobs'/\mu$ to obtain Eqn. 10.  Our 
expressions for number counts agree with Schneider (2006). Note that the 
unlensed number counts are independent of position on the sky, as are 
 $\overline{\mathcal{S}}$ and $\overline{\mathcal{S}^2}$. 
Eqns. 11 and 12 for the observed flux moments 
can be obtained similarly
and easily generalize to the $n$-th moment as 
$\langle \snobs \rangle = 
\langle \mu^n\rangle/\langle \mu \rangle \ \overline{\mathcal{S}^n}$. 
This expression changes if there
is a lower or higher limit to the integral over $S$. For instance 
in the case of an upper limit  $S_{\rm cut}$, we can 
generalize to obtain 
\bea
\snobs(<S_{\rm cut})=\frac{1}{\langle\mu\rangle} \int d\mu\ P(\mu)
\mu^n \ 
\overline{\mathcal{S}^n}\left(<\frac{S_{\rm cut}}{\mu}\right) \,.
\eea

Applications to galaxy clusters are discussed below. Another
application is the contribution of unresolved point sources 
to CMB 
anisotropies, given by $C_{\ell}=\barsvar(<S_{\rm cut})$. 
The upper cutoff $S_{\rm cut}$ is usually introduced to 
remove resolved objects brighter than the cutoff. With lensing the 
observed contribution is enhanced -- given by the above result 
with $n=2$. The enhancement depends on the slope of the $dn/dS$
relation at the cutoff. Finally we note that with 
a flux limit Eqn. 11 is no longer true, since surface 
brightness is only conserved when integrated over all fluxes.

\section{Galaxy Clusters}

Galaxy clusters produce magnifications ranging from $\sim 10$\% 
enhancements above unity to factors of
several or more as one approaches the critical curves. As a result both 
number counts of background galaxies and the flux moments of 
unresolved background sources are significantly altered.   

Consider the unlensed average number of background galaxies
within a cluster solid angle $\Delta \Omegavir$ defined by its virial 
radius, i.e.
$N_{\rm vir}=\Delta\Omegavir \ n(>S)$.
Notice that in the unlensed case $\Delta \Omegavir=\Delta \Omega$, i.e. 
the virial radius of the cluster is actually the intrinsic angle. 
With lensing the virial radius is now the observed solid angle,
i.e. $\Delta \Omegavir=\Delta \Omegaobs$, and we have
\begin{eqnarray}
\langle\Nobs(>\Sobs)\rangle_{\rm vir}
&=&\Delta \Omegavir 
\langle\nobs(>\Sobs)\rangle \\ \nonumber
&=&
\frac{1}{\langle\mu\rangle}\int d\mu P(\mu)\  N\left(>\frac{\Sobs}{\mu}\right)_{\rm vir} \,. \nonumber
\end{eqnarray}
Note that $P(\mu)$ is now the magnification probability within the 
cluster virial radius (in the source plane as before). In the limit of 
constant magnification inside the virial radius we get $\Nobs = N/\mu$. 
Furthermore, if there is no lower limit to the flux integral 
$\Sobs \rightarrow 0$, we have 
$\langle \Nobs \rangle =  N/\langle \mu \rangle$, 
i.e. we observe fewer galaxies in the line of sight of clusters.  
Note that our result for $\Nobs$ appears to differ from some of the
     literature (e.g. Schneider 2006), but the difference is that we
     use the same solid angle for the observed and unlensed case,
     because the only solid angle in town is the observed size of the
     galaxy cluster. 

The intrinsic mean flux and mean square flux 
within the cluster solid angle $\Delta \Omegavir$ are
$
\overline{\mathcal{S}}_{\rm vir} =\Delta \Omegavir\ 
\overline{\mathcal{S}}$ and 
$\overline{\mathcal{S}^2}_{\rm vir} =\Delta \Omegavir \ 
\overline{\mathcal{S}^2}
$. 
With lensing 
we have 
\bea
\langle \sobsbar \rangle_{\rm vir} &=&\Delta \Omegavir \langle \sobsbar \rangle =  \overline{\mathcal{S}}_{\rm vir} \,, \\
\langle \svar \rangle_{\rm vir} &=&\Delta \Omegavir \langle 
\svar\rangle=\frac{\langle \mu^2\rangle}{\langle \mu \rangle}
\overline{\mathcal{S}^2}_{\rm vir}\,.
\label{eqn:SZvariance}
\eea

\begin{figure*}
\resizebox{85mm}{!}{\includegraphics[angle=0]{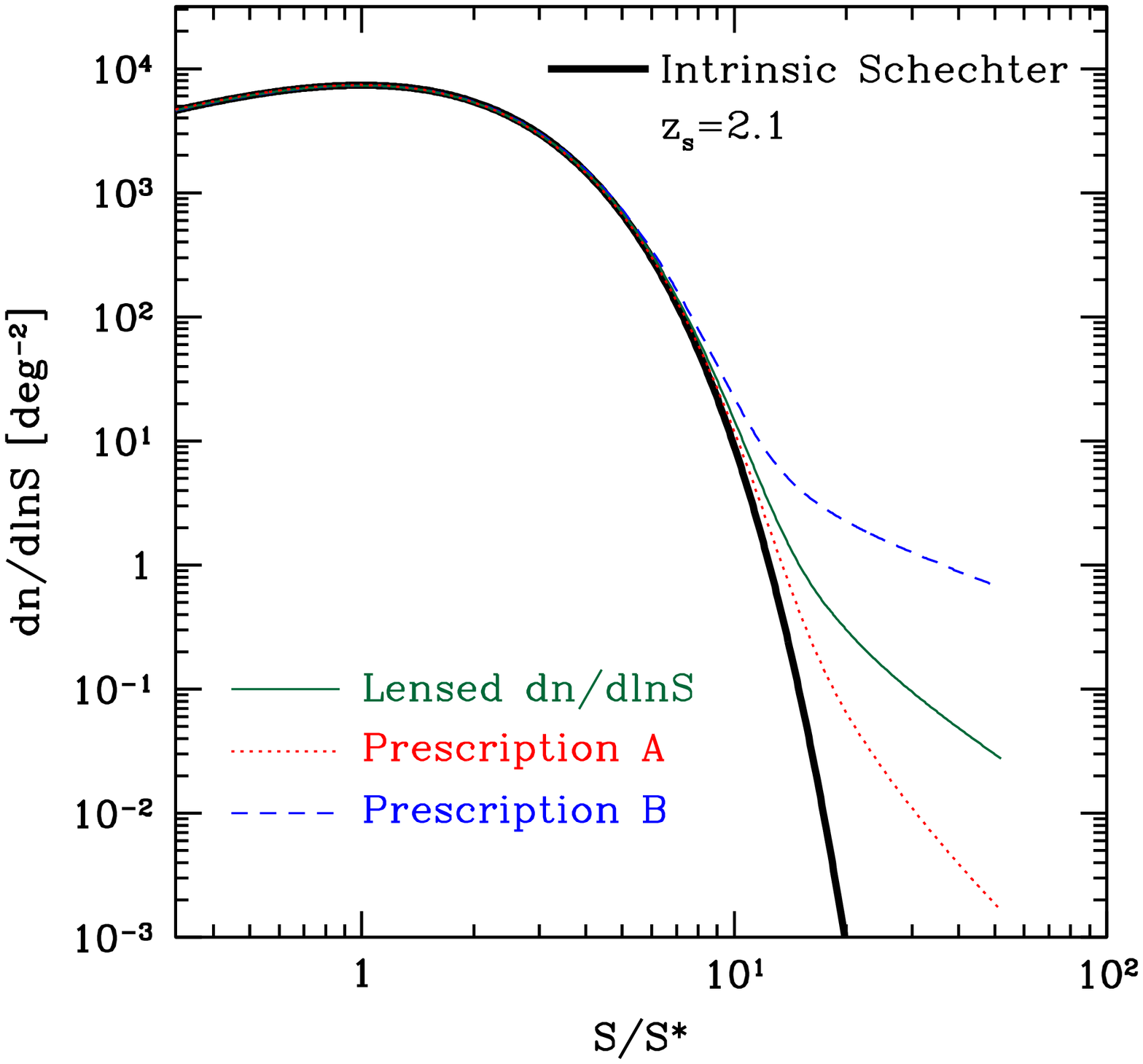}}
\resizebox{85mm}{!}{\includegraphics[angle=0]{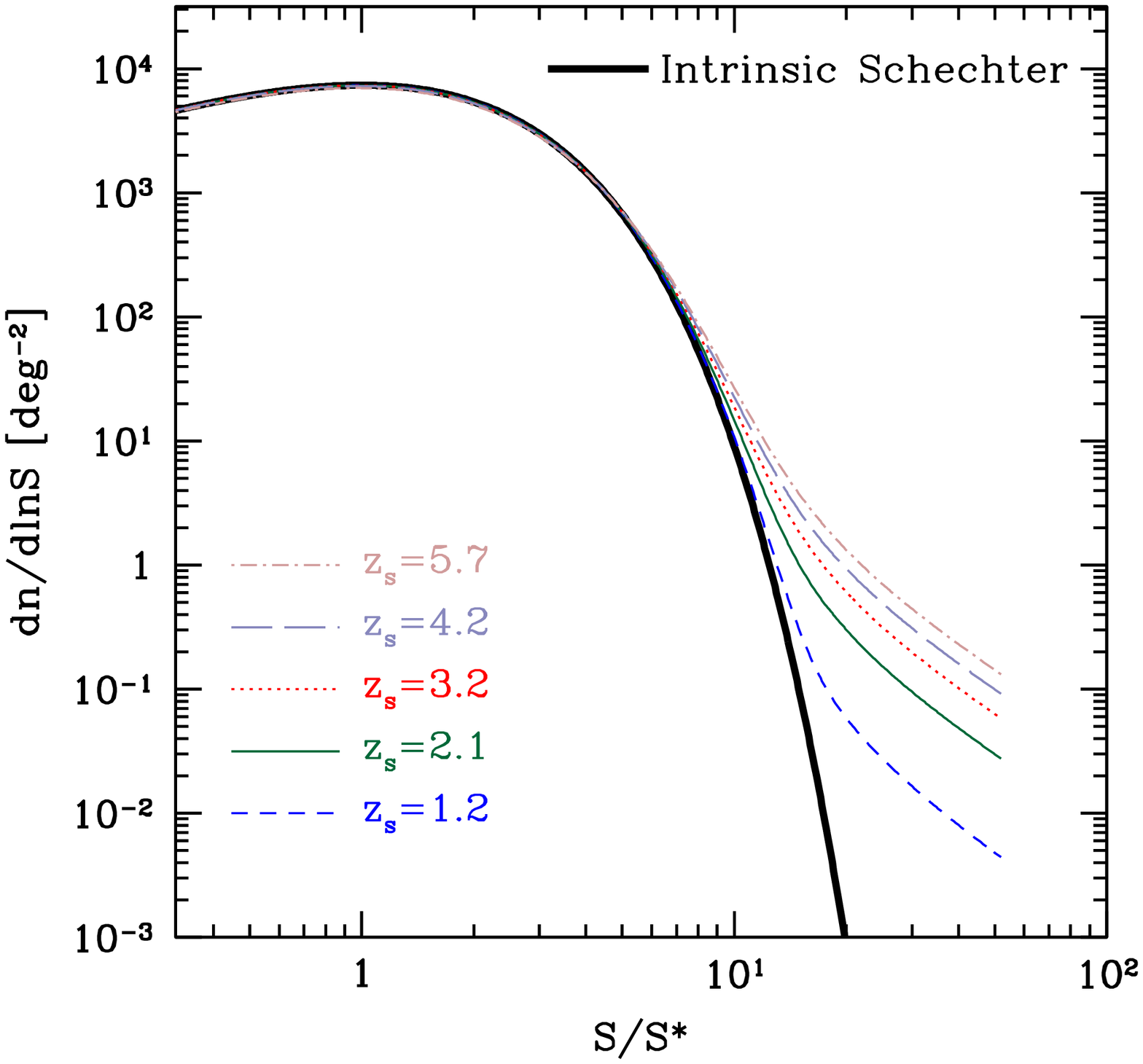}}
  \caption{ {\it Left panel}: 
Intrinsic and lensed galaxy number density ${\rm dn/dlnS}$ 
using different lensing prescriptions.
The true underlying distribution is assumed to be of a Schechter type, 
which is then lensed by intervening halos.  The solid (green) curves use 
Eqn. 9; the two alternatives shown (and described in Section 5.2) 
can change the predicted 
counts by an order of magnitude or more at the high luminosity end. 
{\it Right panel}: As above, but for different source redshifts. 
}
\label{fig:dNdS}
\end{figure*}

Consider a cluster of observed radius 1 arcmin. The average flux
from background galaxies crossing this 1 arcmin cluster is the same
with or without lensing: with lensing, the background galaxies are 
brighter but less numerous and the two effects exactly 
compensate. So, lensing does not create a bias by increasing
the expected flux within the cluster solid angle relative to a 
random 1 arcmin patch of the sky, as long as there is no flux cutoff. 


For SZ surveys, the mean flux from background galaxies is subtracted
in obtaining the SZ decrement or increment (depending on the observed
frequency), but there is additional scatter due to the Poisson error from 
shot noise fluctuations (a given cluster having more or less galaxies than
the expected average). This effect  is important for SZ surveys, where
high-$z$ submillimeter galaxies can contaminate the SZ signal 
(White \& Majumdar 2004; Knox et al. 2004; Lima, Jain \&
Devlin 2009). Lensing enhances 
this source of scatter significantly, as the factor 
$\langle \mu^2\rangle/\langle \mu \rangle$ for galaxy
clusters can be quite large -- see Fig. \ref{fig:avemu}. The estimates
shown are based on an analytical model of cluster halos that uses NFW profiles
and elliptical iso-potential contours (Lima, Jain \& Devlin 2009). 
Our estimates are conservative in that we include cluster ellipticity 
but not substructure. Fig. 2 also indicates the error made in
approximating moments of the flux by using the mean magnification
(e.g. Refregier \& Loeb 1997). 

Notice that if we attempt
to remove galaxies above some flux $S_{\rm cut}$, which
could be removed if they were resolved, 
we introduce
a difference between the observed and intrinsic average flux 
through the cluster, i.e. 
$\langle \sobsbar \rangle \ne \overline{\mathcal{S}} $.
In that case, the mean CMB flux is not equally contaminated by 
background galaxies and subtracting this mean flux from the cluster flux 
does not cancel the galaxy contribution on average. Therefore, 
removing bright galaxies from the sample {\it biases} the SZ signal. 
An observationally relevant situation arises if a flux cutoff is used
to identify sources -- thus altering both the number counts and flux
contribution from unresolved sources (e.g. Refregier \& Loeb 1997).

\section{Illustrative number-magnitude relations}

\subsection{Power law}
It is common in the magnification bias literature to consider the case
of a local power law in the logarithm of the number counts. We then
obtain for the observed cumulative number density:
\begin{eqnarray}
n(>S)\propto S^{-\alpha}\ \rightarrow \ 
\nobs(>\Sobs) = \frac{\langle\mu^\alpha\rangle}{\langle\mu\rangle} \ 
n(>\Sobs) . 
\label{eqn:powerlaw1}
\end{eqnarray}
Working with apparent magnitudes instead of fluxes this gives
(using $m=-2.5 {\rm log}_{10}S + {\rm Constant}$)
\begin{eqnarray}
n(<m)\propto m^{s}\ \rightarrow \ 
\nobs(<m_{\rm obs}) =
\frac{\langle\mu^{2.5 s}\rangle}{\langle\mu\rangle} \ n(<m_{\rm obs}) . 
\label{eqn:powerlaw2}
\end{eqnarray}
Note that the use of the power law for $n(>S)$ allows us to simplify
the integral over $\mu$. The above equation
agrees with the standard expression (e.g. Broadhurst, Taylor \&
Peacock 1995):  $\nobs(<m) = \mu^{2.5s-1}\ n(<m)$ for the case of
constant magnification. For variable magnification, one needs to
evaluate the averages as above. If one simply uses the factor $\langle
\mu^{2.5s-1}\rangle$ behind a galaxy cluster instead of
Eqn. \ref{eqn:powerlaw2}, an error in the number counts can result. 
The error ranges from 3\% for a cluster of mass $10^{14}\ h^{-1}\Msun$ to 6\%
for a mass of  $10^{15}\ h^{-1} \Msun$. This would result in an equivalent bias
in the inferred cluster mass. 
Mass estimates that
rely on number counts may be feasible for large samples of
clusters from future surveys.

To contrast the results from Eqn. 9 with other formulae used in the
literature, consider first the naive generalization to variable
magnification (which is correct in the image plane): 
\begin{eqnarray}
{\rm \bf Prescription\ A:}\ \frac{d\nobs(\Sobs)}{d\Sobs} = 
\int d\mu \frac{P(\mu)}{\mu^2}\
\frac{dn}{dS}\left(\frac{\Sobs}{\mu}\right)\,.
\end{eqnarray}
This underestimates the lensing effect. 
Alternatively some authors drop the $\mu$ factors
altogether (Paciga, Scott \& Chapin 2009), 
which overestimates the lensing effect: 
\begin{eqnarray}
{\rm \bf Prescription\ B:}\ \frac{d\nobs(\Sobs)}{d\Sobs} = 
\int d\mu P(\mu)
\frac{dn}{dS}\left(\frac{\Sobs}{\mu}\right)\,.
\end{eqnarray}

%
%
%

\subsection{Schechter luminosity function}

A single power law number-magnitude relation is lensed into an observed relation with 
same power law (but different amplitude). However if the intrinsic 
$dn/dS$ is not a power law, then lensing
changes the shape of the distribution as well. High magnification events shift galaxies with
low fluxes to high fluxes -- hence if $dn/dS$ falls sharply at high $S$, magnification can 
significantly enhance the counts at these fluxes. 

We illustrate the effect of magnification for realistic galaxy populations by 
considering a Schechter function 
(Schechter 1976):
\begin{eqnarray}
\frac{dn(S)}{dS} &\propto& \left(\frac{S}{S^*}\right)^\alpha 
                    e^{-S/S^*} \,.
\end{eqnarray}
For galaxies observed in a narrow redshift
interval, such a $dn/dS$ relation can arise due to the Schechter
luminosity function of the population.

In Fig.~\ref{fig:dNdS} we show the intrinsic distribution and the 
lensed versions according to the two incorrect prescriptions (denoted A and B) 
mentioned above, as well as the correct prescription of Eqn. 9. 
We assume all 
 galaxies  are at redshift $z_s=2.1$ and use a $P(\mu)$
obtained from N-body simulations by Hilbert et al (2007). We also 
choose $\alpha=0$ in the Schechter function. We plot
 $dn/d\ln S$ to relate more easily to the observational literature
which shows number per unit absolute magnitude. Our results in Fig. 2 may be 
matched with high-$z$ luminosity functions by replacing $S/S^*$ with $L/L^*$.  

The lensing contribution (correctly included in the green
curves) is large for the Schechter function 
at the bright end ($S/S^* >10$). 
Thus for a population of galaxies with a
Schechter luminosity function, the observed $dn/dS$ will not retain 
the exponential tail of the Schechter function. Using 
Eqn. 9 and approximating the Schechter function as having a sharp
cutoff at $S=S^*$, it is easy to see that the integration over $\mu$ 
generates a power law in  $ dn/d\ln S$ whose slope is $-2$, 
due to the asymptotic slope $P(\mu)\propto \mu^{-3}$. The green solid curve 
approaches this slope beyond $S=10 S^*$. 

Choosing a lower (higher) value of $\alpha$ slightly enhances (suppresses) the
lensing contribution at fixed $S/S^*$. Fig.~\ref{fig:dNdS} also shows that the 
difference between the three prescriptions is large at the bright end 
for the Schechter function. 
The $dn/dS$ in Fig. 2 is normalized so that the distribution matches 
submillimeter galaxies 
measured by the Balloon-borne Large Aperture Telescope (BLAST) 
(Devlin et al 2009) at $500$~$\mu$m. 
In a separate study we show that if the sub-mm galaxies lie 
at  $z\sim 2-3$, then lensing of an intrinsic Schechter function can explain 
the observations from BLAST and other surveys (Lima, Jain, Devlin \&
Aguirre 2010). 

Similar comparisons to luminosity function observations in the visible
bands can be 
carried out: it is especially important to include lensing  in the
comparison of low-$z$ data with data at $z\simgt 1$ since the
magnification contribution is significant only at high-$z$. 

\section{Discussion}

We have derived expressions for computing average quantities in 
the observed image plane, given a distribution $P(\mu)$ of magnifications
in the source plane. Our results, summarized in Eqns.  9-12, 
generalize expressions 
found in the lensing literature for the case of constant magnifications.
We illustrated the effect of lensing on steep 
number counts of background galaxies 
and on boosting the contamination that high redshift galaxies induce 
in cluster SZ fluxes. 
The formulae we have presented may be useful for studying the
intrinsic properties of high redshift galaxies and for current and
upcoming cluster surveys. 

The quantitative estimates presented here for galaxy clusters
are based on analytical models of halos. While these incorporate
realistic density profiles as well as halo ellipticity, they 
miss the full complexity of halo bimodality (due to major
mergers) and substructure. These features only enhance the effects of
averaging we have considered. 

Finally we note that the lensing effects discussed here do not impact
the predicted (magnification induced)
cross-correlation of number counts measured in different redshift bins
(Moessner \& Jain 1998). Such cross-correlations depend on the
two-point cross-correlations of magnification with the galaxy
density. Thus measurements of magnification bias from galaxy-quasar 
cross-correlations 
or of its contaminating effect on high-$z$ ISW
cross-correlations 
are unaffected by the spatial averaging issues
discussed here. 

\smallskip

\noindent {\it Acknowledgments:} We thank Anna Cabre, 
Yan-Chuan Cai, Mark Devlin, 
Mike Jarvis and Ravi Sheth
for helpful discussions, and Stefan Hilbert for sharing 
his simulation results. We especially thank Gary Bernstein and 
Peter Schneider for sharing their insights and knowledge of the
history of the field. This work was supported in part by an NSF-PIRE 
grant and AST-0607667.

\end{document}